\documentclass[review]{elsarticle}
\usepackage{booktabs} 
\usepackage{subfigure}
\usepackage{graphicx}
\usepackage{algorithm}
\usepackage{algpseudocode}
\usepackage{amsmath}
\usepackage{url}
\usepackage[misc]{ifsym}
\usepackage{color}

\usepackage{multirow}  %

\usepackage{lineno,hyperref}
\modulolinenumbers[5]

\journal{Journal of Systems and Software}

\begin{document}

\begin{frontmatter}
\title{An Android Application Risk Evaluation Framework Based on Minimum Permission Set Identification}


%
%
\author[mymainaddress]{Jianmao Xiao}
\author[mymainaddress]{Shizhan Chen}
\author[mysecondaryaddress]{Qiang He}
\author[mymainaddress]{Zhiyong Feng}
\author[mymainaddress]{Xiao Xue\corref{mycorrespondingauthor}}
\cortext[mycorrespondingauthor]{Corresponding author}
\ead{jzxuexiao@tju.edu.cn}

\address[mymainaddress]{College of Intelligence and Computing, Tianjin University, Tianjin, China}
\address[mysecondaryaddress]{School of Software and Electrical Engineering,\\
	Swinburne University of Technology, Hawthorn, VIC 3122, Australia}

\begin{abstract}
Android utilizes a security mechanism that requires apps to request permission for accessing sensitive user data, e.g., contacts and SMSs, or certain system features, e.g., camera and Internet access. However, Android apps tend to be overprivileged, i.e., they often request more permissions than necessary. This raises the security problem of overprivilege. To alleviate the overprivilege problem, this paper proposes MPDroid, an approach that combines static analysis and collaborative filtering to identify the minimum permissions for an Android app based on its app description and API usage. Given an app, MPDroid first employs collaborative filtering to identify the initial minimum permissions for the app. Then, through static analysis, the final minimum permissions that an app  really needs are identified. Finally, it evaluates the overprivilege risk by inspecting the app’s extra privileges, i.e., the unnecessary permissions requested by the app. Experiments are conducted on 16,343 popular apps collected from Google Play. The results show that MPDroid outperforms the state-of-the-art approach significantly.
\end{abstract}


\end{frontmatter}


\section{Introduction}
In the past decade, the popularity and ubiquitous use of smartphones have greatly fueled the growth of mobile application (referred to as app hereafter). According to AppBrain\footnote[1]{http://developer.Android.com/guide/platform/index.html.}, as of September 2018, the number of available apps on Google Play-the world largest Android app store, has exceeded 2.8 million. Apps are now playing an extremely important role in our daily life. Many mobile users store a lot of sensitive and private data on their devices. Such data are at the risk of exposure to malicious activities, which has become a major vulnerability of the entire mobile ecosystem~\cite{Experimental}.

Android has long been a major target of malicious apps~\cite{An}. One of its major vulnerabilities is the permission mechanism~\cite{Rpe}. Android's permission mechanism requires apps to request permission for accessing sensitive user data, e.g., contacts and SMSs, or certain system features, e.g., camera and Internet access. Thus, the security of Android heavily depends on the effectiveness of this permission mechanism. A major threat is that a malicious app may furtively request extra permissions for accessing users' sensitive and private data. To minimize this threat, some researchers have designed user-oriented permission prompts to ensure that smartphone users are properly notified of the permissions requested by apps. However, due to the complexity of Android's permission mechanism, most of these efforts have proven to be ineffective~\cite{Apr}~\cite{sok}. The main reason is that most users do not fully understand Android's permissions mechanism. They often simply ignore the prompts and accept apps' requests for permissions without inspecting the prompts~\cite{ApU}. As a result, apps can easily obtain extra permissions, which increase the risks of user privacy leaks. This is referred to as the over privilege problem~\cite{Apd}. A study conducted by Yu et al.~\cite{Rtd} shows that more than 80\% of Android apps are overprivileged. The vulnerability of Android's permission mechanism puts mobile users at the risk of privacy leak in the mobile ecosystem~\cite{Svm}. This has become one of the major threats to the health of mobile ecosystem~\cite{Amd}. This threat is made even more serious by benign apps that are overprivileged by excessively requested permissions~\cite{Apd}~\cite{Pei}.

\par
The mainstream approach for enhancing the Android permission mechanism is to identify over-declared permissions requested by an app ~\cite{Cab}~\cite{TAR}~\cite{AMt}~\cite{ASPG} and recommend reasonable permissions for an app ~\cite{Asb}~\cite{Aao}~\cite{Ahc}. A major and common limitation of these approaches is that the requested permissions are considered as the permissions that the app really uses. However,  this is not always true, especially for malicious apps, they often declare more permissions than they really needs. To address this issue, it is important to identify the minimum permissions, i.e., permissions that are truly needed by an app for the implementation of its {functionalities}. When given the minimum permission for an app, whether it is a malicious or benign application, the over-declared permissions can be identified and pruned without impacting the {functionalities} of the app.
\par
In this paper, we propose Minimum Permission for Android (\textbf{MPDroid}), an approach for Android app risk evaluation based on the identification of minimum permissions. MPDroid identifies the initial minimum permissions for the target app by inspecting the permissions requested by apps that are similar to the target app, following the main idea of collaborative filtering-\emph{if two users (apps) u and v have similar behaviors (functional description), they will act on other items (permissions) similarly}~\cite{EAs}. Then, it obtains the final minimum permissions for the target app by using a functionality point\footnote[2]{The functionality point in this paper refers to functionality topic which is obtained from the description of the app by LDA model}-permission identify method based on API-used code permission and the app declared permission. The major contributions of this paper are as follows:
\par
\begin{itemize}
	\item An over-declared permission identification algorithm is proposed. MPDroid employs the LDA technique and an improved collaborative filtering recommendation algorithm to identify and remove over-declared permission by an app. It then obtains a initial minimum permission set corresponding to the app's description (i.e., declaration {functionalities}).
	\item We employ static analysis to statically parse app related code permissions and analyze the permissions that the app actually calls. In addition, we present a {functionality} point-permission set model to further improve the permission configuration of the apps and obtain the final minimum permission set.
	\item Based on MPDroid, a permission-based risk assessment framework is proposed to detect the risk coefficient for the target app, compared with the state-of-the-art methods, the performance of detecting app risks is improved by 67.5\% for the benign apps. In addition, to enable others to use MPDroid, we have published our source code and dataset on GitHub\footnote[3]{https://github.com/ztxjm123/MPDroid}.
\end{itemize}
\par
The rest of this paper is organized as follows: Section 2 motivates this research. Section 3 presents the  risky app identification process. Section 4 evaluates MPDroid experimentally. Section 5 reviews the  related work and Section 6 concludes the paper.

\section{Motivation}
The permissions needed by an app are often related to its {functionalities}, which can be extracted from the app's description. For example, an application that describes itself as a social networking will likely need permissions related to the mobile device's address book and will need the permission “READ\_CONTACTS”. A number of malware and privacy-invasive applications have been known to declare more permissions than their purported {functionality} warrants [33], which is usually considered to be unreasonable. Take a screen wallpaper app named bollywoodlive for example. We parsed its APK file and found that it actually applied for WAKE\_LOCK, CHANGE\_WIFI\_STATE and RECEIVE\_BOOT\_COMPLETED permissions. These permissions are completely irrelevant to its own functional description which may be harm for the privacy of app users.

\par
An app should not request more permissions than necessary to support its {functionalities}, and the developer should minimize the number of permissions required by apps to reduce the app security risk. This is also recommended by the Android official\footnote[4]{ https://developer.android.com/training/articles/security-tips.html\#\\RequestingPermissions}. However, sometimes the permissions requested by an app deviate significantly from the permissions required by the {functionalities} specified in the app's description ~\cite{Dma}, not only the malicious app, but also for many benign apps~\cite{Aso}, {there are also exist declaration} the unnecessary permissions problems. In this context, the research problem is defined as:
\par
\textbf{\emph{Q1:Given an app $a_{i}$ and its {functional} description information $D F_{i}$, then how to obtain the minimum permission set that the app really needs?}}
\par
\begin{figure*}[!t]
	\centering
	\includegraphics[width=4in]{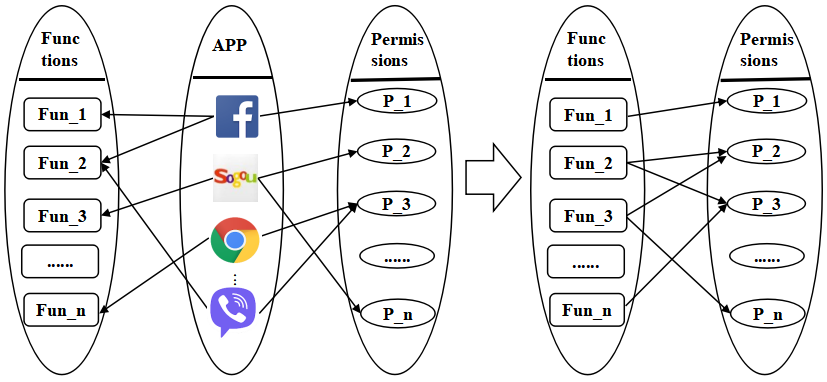}  
	\caption{Association between app's {functionalities} and permissions}
	\label{img1}
\end{figure*}

To solve the above problem, MPDroid is proposed in this paper with the aim to identify the minimum permissions for an app, which is referred to as the \emph{description-minimum permission set} in our work. The basic idea of MPDroid is to establish a mapping relationship between {functionalities} and corresponding permissions to identify abnormal permissions. As shown in Figure 1, the {functionalities} and the permissions of an app are correlated. An app usually provides multiple {functionalities} and requests permissions. A mapping relationship between its {functionalities} and permissions can be established. 
\begin{figure*}[!t]
	\centering
	\includegraphics[width=4in]{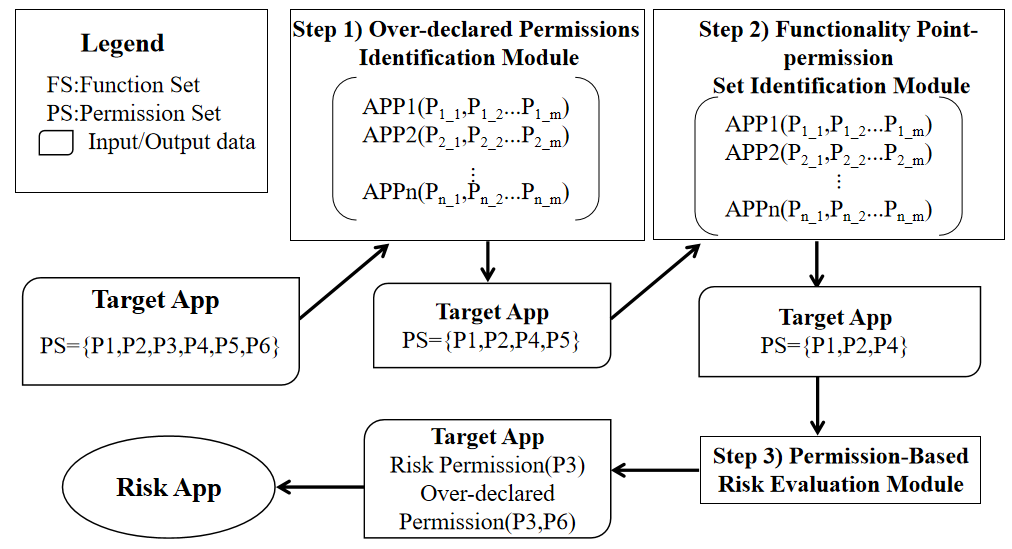} 
	\caption{The Process of Overprivileged App Identification}
	\label{img2}
\end{figure*}

Figure 2 presents MPDroid's process of identifying the minimum permissions of an app. We assume that the target app obtains the permission set PS=$\left\{P1, P2, P3, P4, P5, P6 \right\}$ extracted from its APK. When the target app is processed by the Over-declared Permissions Identification Module (Step 1), the target app's permission set becomes PS=$\left\{P1, P2, P4, P5\right\}$. $P3$ and $P6$ will be removed as over-declared permissions. For instance, Table 1 shows an example of an app processed by MPDroid. The app Bollywood Live applied for 12 permissions. The permission CHANGE\_WIFI\_STATE, GET\_TASKS, RECEIVE\_BOOT\_COMPLETED are over-declared permissions, because the {functionalities} in the application description do not refer to these three permissions.

Meanwhile, the {Functionality} Point-Permission Set Identification Module (Step 2) can further identify the target app's permission set PS=$\left\{P1, P2, P4\right\}$ as its final \emph{description-minimum permission set}. $P5$ is removed because its support degree (details see in Section 3.4) is too low. For example, In Table 1, permission P5 means SET\_WALLPAPER, READ\_LOGS, and SEND\_SMS. Because their permission support degree is lower than the threshold, they will be filtered out. The lower the permission support degree means the lower possible of the permissions required by the app. Then, the rest permissions we regard as the final minimum permissions set of the app.
\newcommand{\tabincell}[2]{\begin{tabular}{@{}#1@{}}#2\end{tabular}}
\begin{table*}[!htbp]
	\centering
	\caption{An Example of App Processed by MPDroid}\label{tab:aStrangeTable}

	\begin{tabular}{|l|p{7.5cm}|}
		
		\hline
		App Name & Bollywood Live \\ \hline
		declared permission & \tabincell{l}{ACCESS\_NETWORK\_STATE, ACCESS\_WIFI\_\\STATE, \textbf{CHANGE\_WIFI\_STATE}, GET\_ACC\\OUNTS, \textbf{GET\_TASKS}, INTERNET, READ\_L\\OGS, \textbf{ RECEIVE\_BOOT\_COMPLETED}, R\\EAD\_PHONE\_STATE, SEND\_SMS, SET\_WALL\\PAPER, WAKE\_LOCK}   \\ \hline
		\tabincell{l}{After processed by\\ step 1} &  \tabincell{l}{ACCESS\_NETWORK\_STATE, ACCESS\_WIFI\_\\STATE,  GET\_ACCOUNTS, 	INTERNET,\\ \textbf{SET\_WALLPAPER}, READ\_PHONE\_STATE,\\ \textbf{SEND\_SMS}, \textbf{READ\_LOGS}, WAKE\_LOCK} 			   \\ \hline
		\tabincell{l}{After processed by\\ step 2} &  \tabincell{l}{ACCESS\_NETWORK\_STATE, ACCESS\_WI\\FI\_STATE, GET\_ACCOUNTS, INTERNET, \\READ\_PHONE\_STATE, WAKE\_LOCK}						 \\ \hline

	\end{tabular}
\end{table*}

Finally, the Permission-Based Risk Evaluation Module (Step 3) identifies the permission $P3$ as a risk permission (defined as equation (14) in Section 3.5). As a result, the target app is regarded as a risk app because $P3$ belongs to both unexpected permissions (defined as equation (13) in Section 3.5) and risk permissions. The details of each module will be discussed in Section 3.

\section{Risk App Identification}
MPDroid employs a permission-based app risk evaluation process for measuring the risk level of an app. Figure 3 shows the process which includes the following 4 phases.
\begin{enumerate}
	\item Over-declared permissions identification. In this phase, MPDroid employs an improved collaborative filtering algorithm to identify and remove over-declared permissions in the app.
	\item Initial Description-minimum Permission Set Identification. In this phase, MPDroid iterates the over-declared permissions identification process to obtain the app's initial \emph{description-minimum permission set}.
	\item {Functionality} Point-permission Set Identification. In this phase, MPDroid recommends the app permissions that combine with the actual declared permission and real requested permissions of the app actually uses by calling APIs. As a consequence, MPDroid further refines the initial description-minimum permission set of the app to obtain the final \emph{description-minimum permission set}.
	\item Permission-based Risk Evaluation. In this phase, MPDroid calculates the risk level of the app and classifies the app as risky or not.
\end{enumerate}
\begin{figure*}[!t]
	\centering
	\includegraphics[width=4.5in]{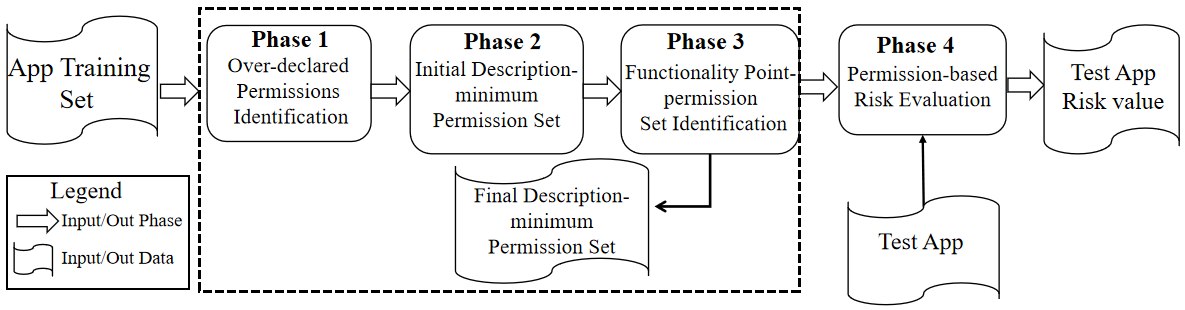}  
	\caption{The Overview of MPDroid}
	\label{img3}
\end{figure*}

\subsection{Definitions}
In order to establish the mapping relationship between app {functionalities} and permissions, MPDroid first identifies the {functionality} points implemented by all the local apps' descriptions, and then establishes a mapping relationship between the app and its {functionalities}. This is referred to as description {functionality} point extraction. Secondly, MPDroid detects the list of permissions that the app actually requests, and establishes a mapping relationship between the app and its permissions. This is referred as declared permission extraction. Here, we give some formal definitions.
\par
\textbf{\emph{Definition 1: An app $a_{i}$ is defined as:}}
\begin{equation}
a_{i}=<DF_{i}, DP_{i}, CP_{i}, \operatorname{Min}P_{i}>
\end{equation}
where $DF_{i}$ represents $a_{i}'s$ {functionalities}, $DP_{i}$ represents  $a_{i}'s$ declared permissions,  $CP_{i}$ represents  $a_{i}'s$ API-based permissions, i.e., permissions parsed from the code, and $MinP_{i}$  represents $a_{i}'s$ minimum permissions identified by MPDroid.
\par
\textbf{\emph{Definition 2: An app's declared {functionalities} are defined as:}}
\begin{equation}
D F_{i}=<D F_{i, 1}, \ldots, D F_{i, k}>, 0<D F_{i, k}<1
\end{equation}
where $DF_{i, k}$  is the probability of declared {functionality} point of the app and $k$ is the number of description {functionality} points {contained in app's description}.
\par
\textbf{\emph{Definition 3: Declarative permission information for Android app.}}
\begin{equation}
D P_{i}=<D P_{i, 1}, \ldots, D P_{i, m}>
\end{equation}
$DP_{i}$ represents the permission set declared by $a_{i}$, and $m$ represents the number of permissions extracted from  $a_{i}'s$ AndroidManifest file.

\par 
\textbf{\emph{Definition 4: Code permission information for Android apps.}}
\begin{equation}
C P_{i}=<C P_{i, 1}, \ldots, C P_{i, n}>
\end{equation}
where $CP_{i}$  represents the code permissions of $a_{i}$. MPDroid first employ static analysis to obtain the Android-related API that the APK calls, followed by the Android-related API to map to the its permissions. Here, $n$ is the number of permissions that are parsed based on the code and $n$$\geq$ 0.

\par
\textbf{\emph{Definition 5: Minimum permission set information for Android app.}}
\begin{equation}
\operatorname{Min} P_{i}=<\operatorname{Min} P_{i, 1}, \ldots, \operatorname{Min} P_{i, q}>
\end{equation}
where $MinP_{i}$ represents $a_{i}'s$ minimum permissions identified by MPDroid, and $q$ is the number of permissions in ${Min} P_{i}$.
\par
Based on the above definitions, MPDroid can now perform data (i.e., {functionality} point and permission data) feature extraction, MPDroid extracts the {functionality} information and the declared permissions from the textual functional description of the Android app and its corresponding APK file respectively. The goal here is to extract the app data features for building the mapping relation between the declared {functionalities} and declared permissions for each app. It consists of the following three contents.
\par
\textbf{Description {functionality} Point Extraction.} We obtain the declared functionality topics for all the app based on their {functionality} descriptions, and use the Latent Dirichlet Allocation (LDA) on the descriptions to cluster app into different {functionality} topics. We define the {functionality} vector set for each app as a $Func$ matrix, formulated as follows:
\begin{equation}
Func=\left[ \begin{array}{cccc}{D F_{1,1}} & {D F_{1,2}} & {\dots} & {D F_{1, n}} \\ {D F_{2,1}} & {D F_{2,2}} & {\dots} & {D F_{2, n}} \\ {\ldots} & {\ldots} & {\ldots} & {\ldots} \\ {D F_{m, 1}} & {D F_{m, 2}} & {\dots} & {D F_{m, n}}\end{array}\right]
\end{equation}
$D F_{i, j}$ represents the probability that app $a_{i}$  belongs to topic $F_{j}$, $0<D F_{i, j}<1$ , $n$ represents the number of topics in all app, $m$ represents the number of apps. For the newly target app, we can get the declaration {functionality} vector of the app by matching the described information with the LDA trained model.
\par
\textbf{Declared Permission Extraction.} For each app,  MPDroid extracts its APK file with apktool\footnote[5]{http://ibotpeaches.github.io/apktool} and obtains the declared permissions from its AndroidMainfest file. By parsing the AndroidMainfest file, the full permission set can be obtained as follows:
\begin{equation}
D P=\left[ \begin{array}{cccc}{D P_{1,1}} & {D P_{1,2}} & {\dots} & {D P_{1, n}} \\ {D P_{2,1}} & {D P_{2,2}} & {\dots} & {D P_{2, n}} \\ {\ldots} & {\ldots} & {\ldots} & {\ldots} \\ {D P_{m, 1}} & {D P_{m, 2}} & {\dots} & {D P_{m, n}}\end{array}\right]
\end{equation}
where  $D P_{i, j}=1$  represents that  $a_{i}$  applies for permission $D P_{j}$ or 0 otherwise.

\par
\textbf{API-based Permission Extraction.}
Certain permissions will be required when an app calls the Android APIs. To find out the real requested permissions that an app actually uses by calling APIs, MPDroid adopts an open-source tool named Androguard\footnote[6]{Androguard: https://code.google.com/p/androguard} to statically analyze the app's APK, and obtains the code permissions of the app. MPDroid first traverses all the code files in the APK, and detects the API in the file to obtain all the Android-related methods. Then, according to the result of Pscout~\cite{Pscout}, the correspondence relation table between the API and the permissions is built, and the Android API obtained by the traversal scan is mapped with the permissions. Finally, we obtain the app’s code permissions. In our study, 16,343 apps were parsed, and a tree-shaped relationship diagram of app API-permission information is constructed. 

\par
Compared to the information extracted from the AndroidManifest file, the permissions obtained from the code are often more accurate, and in addition, the code permissions are the foundation for the implementation of the {Functionality} Point-Permission Set Identification in Section 3.4.

\subsection{ Over-declared permissions identification}
The purpose of this phase is to identify and remove the permissions that are over-declared in the app by the recommendation algorithm for the target app. Figure 4 shows the overall framework of the over-declared permission identification in MPDroid.
\begin{figure*}[!t]
	\centering
	\includegraphics[width=4.5in]{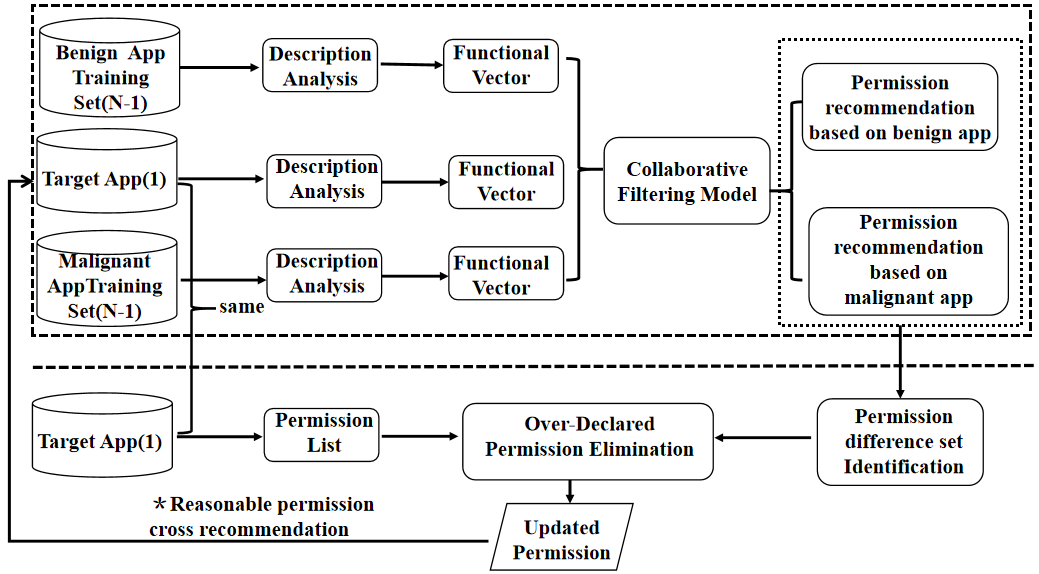} 
	\caption{Overall framework of over-declared permission identification}
	\label{img4}
\end{figure*}
\par
MPDroid leverages two datasets, i.e., the benign app dataset and the malicious app dataset. We divided the benign app dataset into $N$ parts, $N$-1 parts form the benign training set, and the remaining part as the target app set that needs to remove the over-declared permissions. For the malignant app, we perform the same operation. Our goal is to map the declared {functionality} information to the corresponding reasonable permissions for the target app. MPDroid employs an improved collaborative filtering algorithm to recommend permissions for the remaining app set, i.e., the target app. The permission recommendation procedures include following contents.

\subsubsection{Similarity Computation}
We assume that there are $m$ apps. Each app has $n$ {functionalities}, the relationship between apps and {functionalities} is denoted by an $m$$\times$$n$  matrix, i.e., the $Func$ matrix. Each entry $D F_{i j}$ in the matrix represents the probability that the app belongs to this {functionality} point. The larger the probability value, the more likely the app belongs to this {functionality}. Conversely, the smaller the probability value, the less likely the app includes this {functionality}. Here, the ``Euclidean distance" is employed for the similarity computation. ``Euclidean distance" employs the Equation (8) to compute the distance between the target app and the app located in the training set:
 
\begin{equation}
\operatorname Dist(X, Y)=\sqrt{\sum_{i=1}^{n}\left(X_{i}-Y_{i}\right)}
\end{equation}
where X and Y represent the {functionality} vectors of app $a_{i}$  and $a_{j}$ respectively. $\operatorname Dist(X, Y)$ represents the distance between app $a_{i}$  and $a_{j}$, the similarity between app is calculated as equation (9):
\begin{equation}
\operatorname Sim\left(a_{i}, a_{j}\right)=\frac{1}{1+\operatorname Dist (X, Y)}
\end{equation}

\subsubsection{Similar App Selection}
After calculating the similarity values between the target app and the app located in the training set, a set of similar apps can be identified. On the other hand, in the process of permission identification, if a part of the high similarity application is used for recommendation, the effect will be better than the recommendation using all the applications. It is worth noting that since the number of similar members obtained by the Top-K algorithm is fixed each time, in fact, the number of similar members of the app is not well determined and the number of similar members for each app is not necessarily the same, which may lead the quality of similar members of the app being unguaranteed, and resulting in poor recommendation results. Thus, MPDroid employs the similarity threshold method to obtain a subset of similar app by setting a certain threshold parameter T.

\subsubsection{Permission Recommendation}
MPDroid selects similar app with similarity greater than the threshold T as candidate recommendation members for each target app, and then it uses the similarity of the declaration {functionality} vectors of these similar members to perform a comprehensive weighted calculation on the permissions they declare, finally, it obtains a list of the permission recommendation results for the target app. For example, the permission $p_{i}$ that the app $a_{i}$ may declare is weighted according to the similarity value to calculate the recommended permission for the target app, it is similar to the user-based collaborative filtering method. The comprehensive recommendation value is defined as $\mathrm{Rv}_{a_{i}, i}$, and the calculation formula is as equation (10):
\begin{equation}
R v_{a_{i}, i}=k \sum_{a_{j} \in F} \operatorname Sim\left(a_{i}, a_{j}\right) R v_{a_{j}, i}
\end{equation}
where $k$ is a normalizing factor defined as $1/ \sum_{a_{j} \in F} \operatorname Sim \left(a_{i}, a_{j}\right)$ , $F$ refers to the app recommended member set whose similarity is greater than the threshold T, $R v_{a_{j}, i}$ is equals to 1 if app $a_{j}$ declares the permission, otherwise $R v_{a_{j}, i}$ is equals to 0 that indicates app $a_{j}$ does not declare the permission. After calculating a recommended value vector of the target app, we can get a list of recommended permissions ranking from high recommended value to low. The larger the value is, the higher possibility that the permission is needed for the app. Finally, MPDroid uses the adaptive parameter based method [16] to generate the final permission list.

\par
It is worth noting that in order to avoid the wrong removal of normal permissions, MPDroid uses the malicious apps and the benign apps as the training sets separately, and treats the difference between the recommendation result of the malicious app and the benign app recommendation result as the permission difference set. If the permissions are located in the difference set and also locted in the target app's permissions, so we call them the over-declared permissions. Thus for the given target app, the difference set of the two recommended permission is removed from the declared permissions. As the recommendation result of benign app often represents the permissions necessary for the app to implement the {functionality}, so we do not remove this part permissions. But for the results of malicious app recommend permissions often tend to be more, because in addition to the permissions required to support normal {functionalities}, there are also include many dangerous permissions. So we believe that the different set of recommendation permissions between a malicious app and a benign app also represents dangerous permissions.
\par
For each app in the training set, MPDroid extracts the declared {functionality} from the description of the app and then establishs a mapping between {functionalities} and permissions. Note that our method is a reasonable permission cross recommendation process as shown in Figure 4, after updating a set of target app permissions, we put the set of apps back into training set and continue to train until all apps in the training set are recommended with reasonable permissions, i.e., removing the over-declared permissions, compared with the Figure 2 in Section 2, we remove the over-declared permissions ($P3$, $P6$) of the target app through the over-declared permission identification module.

\subsection{Initial Description-minimum permission set identification}
In order to generate a minimum set of permissions corresponding to the description information, we propose an iterative algorithm called description-minimum permission set identification, its purpose is to generate a minimum permission sets corresponding to the Android app description (declaration {functionality}). The whole process of iterations can be described as follow:
\begin{enumerate}
	\item Initialize the training set: We divide the entire benign app set into $N$ parts marked as $<A_{1}, A_{2}, \dots, A_{N}>$. Next we use the rest of $N$-1 parts $<A_{2}, A_{3}, \ldots, A_{N}>$ to identify the over-declared permissions for the target app set $<A_{1}>$ by using the over-declared permission identification method in Section 3.2.
	\item Get the over-declared permissions for $<A_{1}>$ and then remove them.
	\item Update the training set: Put $<A_{1}>$ that had removed the over-declared permissions back to the training set and use the next part of the app as the target app to identify the over-declared permissions. For example, we use the app $<A_{1}, A_{3}, \ldots, A_{N}>$ to identify the over-declared permissions for the $<A_{2}>$. Finally, all the $N$ parts would be removed the over-declared permissions.
	\item After all the $N$ parts have been removed the over-declared permissions, we get the new update dataset, we call this process \textbf{\emph{one-time iteration}}. Next, we repeat the step 1) to step 3) until the permissions do not change. We finally get the minimum permission set of the benign app after several times iteration.
\end{enumerate}

\subsection{{Functionality} Point-Permission Set Identification}
MPDroid obtains the initial minimum permission set corresponding to the target app by Description-minimum permission set identification algorithm in Section 3.3. However, that is only the theoretical result. We need to further combine the permissions of the app actually calls to further refine the minimum permission set corresponding to the target app. So in this phase, we will mine the permission set corresponding to the {functionality} point obtained by the LDA topic model from the perspective of the {functionality} point-permission set. We use the static parsing permissions and the permission of the app declare itself to build a model to obtain each {functionality} point corresponding permissions. Thus, it can further refine the initial \emph{description-minimum permission set}. Through this, we can obtain our final \emph{description-minimum permission set}.

\par
We select all the apps with the same {functionality} topic in $Func$ matrix and traverse all the apps for each topic $T_{m}$. For each app $a_{i}$, the permission that it contains is expressed as $a_{i}=\left[P_{1}, P_{2}, P_{l},. . . P_{N}\right]$. Accordingly, we define the support degree corresponding to each permission as equation (11):
\begin{equation}
R P_{l}\left(a_{i} | P_{l}\right)=P r_{m} \times 1
\end{equation}
where $R P_{l}\left(a_{i} | P_{l}\right)$ represents the support degree for each permission, $P r_{m}$ is the probability corresponding to the {functionality} $T_{m}$.
\par
Next, we add up the values of the same permissions for all apps (assuming that total number of apps is $n$) under each topic $T_{m}$. Then for each permission $P_{l}$, its total permission support degree to the topic $T_{m}$ is:

\begin{equation}
R P_{l}\left(T_{m} | P_{l}\right)=\frac{\sum_{i=1}^{n} R P_{l}\left(a_{i} | P_{l}\right)}{\sum_{i=1}^{n} \operatorname{Pr}_{m}\left(a_{i} | T_{m}\right)}
\end{equation}

where $\operatorname{Pr}_{m}\left(a_{i} | T_{m}\right)$ represents the probability that $a_{i}$ belongs to $T_{m}$. We calculate the permission with the highest $m$ value as the most relevant permission for the topic. If give out a new app with the {functionality} topic is $a_{j}=\left[T_{1}, T_{2}, T_{m}, \ldots, T_{N}\right]$, and its corresponding permissions probability is $\left[ {Pr}_{1}, {Pr}_{2}, {Pr}_{m}, \ldots, {Pr}_{N}\right]$. As for the topic $T_{m}$, the corresponding most relevant $t$ permissions value is $\left\{R P_{1} \times {Pr}_{m}, R P_{2} \times {Pr}_{m}, \ldots, R P_{t} \times {Pr}_{m}\right\}$. Then we plus the permission values of all the same permissions in $\left[T_{1}, T_{2}, T_{t}, \ldots, T_{N}\right]$ and arrange them from large to small to obtain the permission that the $R P_{l}$ value is greater than the value of support degree threshold $\theta_{\text {support}}$, i.e., the permissions  recommended according to the {functionality} point directly.
\par
Finally, we inspect whether the initial \emph{description-minimum permission set} in Section 3.3 exists the permission that the support degree is too low. If it exists, we remove these permissions. This way, we obtain the final \emph{description-minimum permission set} according to the target app. As show in Figure 2, the permission set of the target app changes from $\left\{P1, P2, P4, P5\right\}$ to $\left\{P1, P2, P4\right\}$ in this phase since $P5$ is the permission whose support degree is too low.

\subsection{Permission-Based Risk Evaluation}

We can perform permission-based risk evaluation with MPDroid. Figure 5 shows a permission-based risk evaluation framework. The key idea is to take use of the \emph{description-minimum permission set}, when a target app comes, we recommend the permissions to the target app by use the collaborative filtering method. Then, we compare the difference between the recommendation permissions and the actually declared permissions of the target app, and through this, we can further calculate the risk value for the target app.

\begin{figure*}[!t]
	\centering
	\includegraphics[width=4.5in]{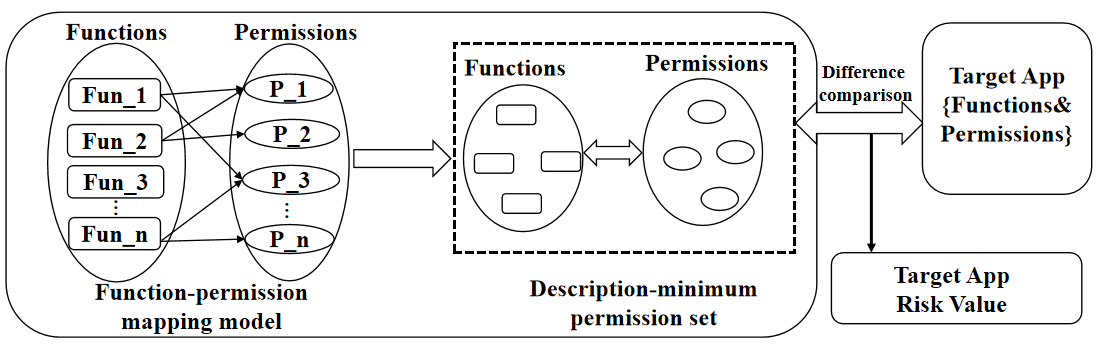} 
    \caption{Permission-based risk assessment framework}
	\label{img5}
\end{figure*}

Considering the fact that almost all the malicious apps have declared more permissions than what the original application needs, we calculate the risk level based on the gap between the malicious app and the benign app. Specially, the minimum permission set of the benign app has been identified already. Since the permissions recommended by benign apps often represent normal and required permissions, so the one that is not recommended in the app's declare permissions can be regarded as abnormal or does not support the permissions required by its declare {functionalities}. Therefore, given a target app $a_{i}$, its declared permissions are $D P_{i}=<D P_{i, 1}, \ldots D P_{i, M}>$ and the recommended candidate permissions are $R P^{*}\left(a_{i}\right)=<p_{i, 1},p_{i, 2}, \dots ,p_{i, N}>$. We can define the unexpected permissions $U P\left(a_{i}\right)$ as equation (13):
\begin{equation}
p_{i} \in U P^{*}\left(a_{i}\right) \leftrightarrow p_{i} \in D P\left(a_{i}\right)-R P^{*}\left(a_{i}\right) \cap D P\left(a_{i}\right)
\end{equation}
where $* \in\{B, M\}$. $B$ represents the permissions recommended by the benign app, $M$ represents the list of permissions recommended by the malicious app.
\par
Also, the recommended permissions by the malicious set $R P^{M}\left(a_{i}\right)$ will contain not only the necessary permissions but also the risk permissions. On the contrary, most of the permissions recommended by the benign set are likely to be necessary. Thus, we define the risk permissions based on the permission gap $R P^{M}\left(a_{i}\right)-R P^{M}\left(a_{i}\right) \cap R P^{B}\left(a_{i}\right)$. The gap between the malicious sets and benign sets can be used to find the risk permissions, such as the permission $P3$ of target app in Figure 2. Here, the risk permissions $R i P\left(a_{i}\right)$ can be formally defined as equation(14):
\begin{equation}
p_{i} \in RiP\left(a_{i}\right) \leftrightarrow p_{i} \in R P^{M}\left(a_{i}\right)-R P^{M}\left(a_{i}\right) \cap R P^{B}\left(a_{i}\right)
\end{equation}
\par
If the target app has an unexpected permission and also belongs to the risk permissions, we consider the app is a risky app, a risky app is defined as follow equation:

\begin{equation}
a_{i} \in R i s k \leftrightarrow p_{i} \in R i P\left(a_{i}\right) \& p_{i} \in U P^{B}\left(a_{i}\right)
\end{equation}
\par
Finally, we can calculate the risk value as follow equation:
\begin{equation}
{Risk}\left(a_{i}\right)=\sum_{p_{j} \in\left(U P^{\mathrm{B}}\left(a_{i}\right) \cap R i P\left(a_{i}\right)\right)} r\left(p_{j}\right)
\end{equation}
\par
We calculate the risk level based on the permissions protection levels: normal and dangerous according to the Android permission mechanism. The scores of the permissions for two protection levels are assigned as 1 and 2 respectively. Here, $r\left(p_{j}\right)$ refers to the risk of  permission based on its protection level.

\section{Experiments}
\subsection{Experimental Setup}
We use the app market dataset from~\cite{Ams}. After processing the dataset, we choose 16,343 app with at least 100+ downloads and five stars in Google Play as the benign dataset finally. The malicious dataset was retrieved from VirusShare\footnote[7]{http://virusshare.com}. Since the VirusShare does not offer app descriptions, so we use the package identifier to map it into the one in the ``app market" dataset. Finally, we find 524 matched items and regard them as the malicious app dataset. For the dataset of {functionality} point-permission set, we have selected 32,671 apps as the dataset to mine {functionality} point-permission relationships, which contains both the text information and APK files. 
\par
There are two kinds of permissions in the Android ecosystem: one is the system permission which is defined by the Android platform, and the other is the custom permission defined by developers themselves. In our experiment, we do not consider the custom permissions in the app because they are only defined and used by developer themselves. Especially, as the Android platform has many different versions, we take all permissions that have ever been defined, whatever used or not. Table 2 summarizes the dataset~\cite{Doi} and Table 3 summarizes the experiment parameters.

\begin{table}[!htbp]
	\centering
	\caption{DataSet Summary}\label{tab:aStrangeTable}
	\setlength{\tabcolsep}{1.6mm}{
		\begin{tabular}{|c|c|}
			\hline
			App Datasets &Number of Applications \\  
			\hline
			Benign App &16343 \\  
			\hline
			Malicious App &524 \\  
			\hline
			\tabincell{c}{\ App for {Functionality} Point-Permission \\Set Identification}  &32671 \\  
			\hline
			Permission Dataset &Number of Permissionss \\  
			\hline
			System permissions&285 \\ 
			\hline
		\end{tabular}
	}
\end{table}
\begin{table}[!htbp]
	\centering
	\caption{Experimental Parameter Descriptions}\label{tab:aStrangeTable}
	\setlength{\tabcolsep}{1.6mm}{
		\begin{tabular}[t]{|c|l|}
			\hline
			Symbols &Descriptions \\  
			\hline
			Topic Number &the numbers of the topic in LDA \\  
			\hline
			Similarity threshold ${T_b}$, ${T_m}$ &\tabincell{l}{the threshold value of the similarity\\ member selected employed for \\permission recommendation} \\
			\hline
			Support degree threshold  $\theta_{\text {support}}$  &\tabincell{l}{the support degree filter value in \\{Functionality} Point-Permission set \\identification phase}\\
			\hline
			Test set ratio & the ratio of the test app set \\  
			\hline
		\end{tabular}
	}
\end{table}
\subsection{Analysis the results for each phase}
\subsubsection{Over-declared permissions identification}
Over-declaration permission identification process has been described in Section 3.2. In fact, the results by the LDA model often result in a sparse declaration {functionality} matrix. Therefore, in the specific implementation process, we made an inverted table according to the {functionality} relevance, and only the {functionality}-related app is calculated, which can greatly reduce the amount of calculation. On the other hand, considering that the number of benign app is much larger than the malicious app, so we select the threshold for the benign app as $T_{b}$=0.6 and malicious app we have chosen the threshold $T_{m}$=0.4 according to~\cite{Asb}. Furthermore, we use the method based on adaptive parameter in~\cite{Asb} to filter out all the permissions that are significantly higher than other permissions until the data is balanced. Finally, we remove the difference set recommended by the malicious and benign app in declared permissions, thus an over-declared permission identification process is completed.

\subsubsection{Initial Description-minimum permission set identification}

In this phase, our goal is to identify the minimum permission set information corresponding to the app. We first randomly divide the benign apps into 5 parts, i.e., $N$=5 in Section 3.3. Then we perform the initial \emph{description-minimum permission set} identification method. In the actual experiments, we found that after 2 iterations, the permission data no longer changes, which means that the initial \emph{description-minimum permission set} is obtained.
%

\begin{table*}[!htbp]
	\centering
	\caption{An Example of the Initial Description-Minimum Permission Set}\label{tab:aStrangeTable}
	
	\begin{tabular}{|l|l|l|}
		\hline
		Functional Description & \tabincell{c}{Over-Declared \\ Permission} & \tabincell{c}{Initial Minimum \\ Permission Set} \\  
		\hline
		\multirow{10}{*}{\tabincell{l}{mission real music studio stude\\nt love reach locat midtown stu\\dio offer instruct level excel mu\\sicianship piano voic drum invo\\lv north food spring station ped\\estrian access teacher cost lesso\\n registr date event program ap\\p latest watch tab wall share fa\\mily class amp loop tube class \\channel latest music video}
		} 
		& & \tabincell{l}{ACCESS\_COARSE\\\_LOCATION} \\
		&CALL\_PHONE &\tabincell{l}{ACCESS\_FINE\_LO\\CATION} \\
		& AMERA & \tabincell{l}{ACCESS\_NETWORK\\\_STATE}\\
		& SEND\_SMSC&RECORD\_AUDIO \\
		& &\tabincell{l}{ WRITE\_EXTERNAL\\\_STORAGE}\\
		& &\tabincell{l}{READ\_PHONE\_STA\\TE}\\
		
		\hline 
	\end{tabular}
\end{table*}
\par
In addition, to illustrate our experimental results, we also provide an example of the minimum permission set identification as shown in Table 4. From that, we can see that an app in our benign dataset requires 10 permissions which are more than it actually needs. We identify the over-declared permissions and generate the initial minimum permission set. The words like ``locat, video, music, north" in the {functionality} descriptions are related to `location' or `audio'. Thus, these permissions like RECORD\_AUDIO and ACCESS\_FINE\_LOCATION enable the basic {functionalities} of the app and they are in the minimum permission set identified by MPDroid. On the contrast, we identify three permissions which are over-declared from the app descriptions: the functional descriptions do not include words related to `call', `sms' or `camera'. This means that these three permissions are unrelated permissions different from the most apps with the similar description, which may lead to the leak of sensitive information.

In fact, the apps with a set of fixed {functionalities} will contain a minimum set of permissions corresponding to itself in theory. However, in the actual operation process, due to the inaccuracy of the app description data, the limitation of the amount of the app data, and the accuracy of the recommendation algorithm, it is impossible to find the minimum permission set. Through our method, we can continue to narrow the scope of the declare permissions without removing the reasonable permissions. This way, we can obtain we can get the most reasonable permission set of the app (i.e., the minimum permission set in this article). According to our iterative algorithm in Section 3.3, it does not delete reasonable permissions. The specific theoretical proof is as follows:
	
For a testing app, suppose its declared permission set is $D P$, and the permission set recommended by the bengin app is $B P$, and the permission set recommended by the maclicious app is $M P$, assume the minimum permissions set of the app is $MinP$. From a overall perspective: MPDroid will remove some permissions for each iteration. Thus, the number of permissions originally declared by the app will gradually decrease. However, since $MinP \subseteq B P, MinP \subseteq M P$, there is $MP-BP \cap MinP=\emptyset$, i.e., the permissions that are removed each time are not the permissions in the minimum permission set, which can guarantee the lower bound of the final recommended result is greater than or equal to $MinP$.
	
From the perspective of the recommendation process: the difference set of $M P-B P$ will removed from the declaration permission set $D P$ in each iteration, and $M P$ includes a minimum set permissions and the risk permissions distinct from $B P$, i.e., the result of $M P-B P$ is risk permission (also called over-declared permission). In the $n$-th iteration, the recommendation result of the $M P$ is unchanged, and $B P$ removes some of the $M P$ recommended permissions in the $n$-1 iterations. Thus, the permissions are gradually reduced compared to $n$-1 times, which also leads to fewer permissions being removed in $D P$. When the permission gap is reduced to 0 or close to 0, the permissions removed during the iteration process will be fewer, and the $D P$ set tends to be stabilize. At this time, the $D P$ set basically does not include the risk permissions and basically converges. Therefore, MPDroid does not incorrectly remove the permissions that the app actually uses.

In addition, the over-declared permission identification of the testing set app is based on difference set recommended by the malicious training set apps and the benign training set apps, as shown in phase 1 and phase 2 in Figure 3. In this step, we mainly identify the over-declared permission for 80\% of benign data (16,343) in training set, i.e., 13,075 apps, and the result is that 635 permissions were removed, involving 479 apps, accounting for 3.66\% of the total apps. For the bengin test apps (3,268), the result is that 301 permissions were removed, involving 205 apps, accounting for 6.27\% of the total benign testing apps. This also shows that there are also cases of over-declared permissions even for the benign app. However, for the malicious testing apps, 58 permissions were removed, involving 37 apps, accounting for 35.58\% of the total malicious testing apps. These results show that the proportion of malicious apps which with over-declared permissions is much higher than that of a benign apps, which is in line with the real-world common sense.

\subsubsection{{Functionality} Point-Permission Set Identification}
The purpose of this phase is to mine the relationship between the {functionality} points obtained by LDA and the declared permissions directly. We calculate the relationship between app {functionality} points and permissions in real calls, and get the closest permission for each {functionality} point, then we get the actual permission set for each app. During the experiment, 32671 benign apps were selected, which contains the text information and APK file.
\par
We map the {functionality} point information and the declared permission information of all apps, and implement the {functionality} point permission set mining according to Section 3.4. Then we obtain the support degree of each {functionality} point corresponding to the permission and arrange them from high to low. In terms of the code permission, we do the same thing. Then, we select the permission with the support degree greater than 0.1 and take the union of code permission and the declaration permission corresponding to the same {functionality} point. Meanwhile, we combine the results of the two to obtain the {functionality} point-permission set relationship. Finally, we inspect whether the initial minimum description permission set after the iterations in Section 3.3 exist the low support degree permissions. If it exists, remove these permissions, and thus we get the final description-minimum permission set corresponding to the app.
\par
\begin{figure*}[!t]
	\centering
	\includegraphics[width=5in]{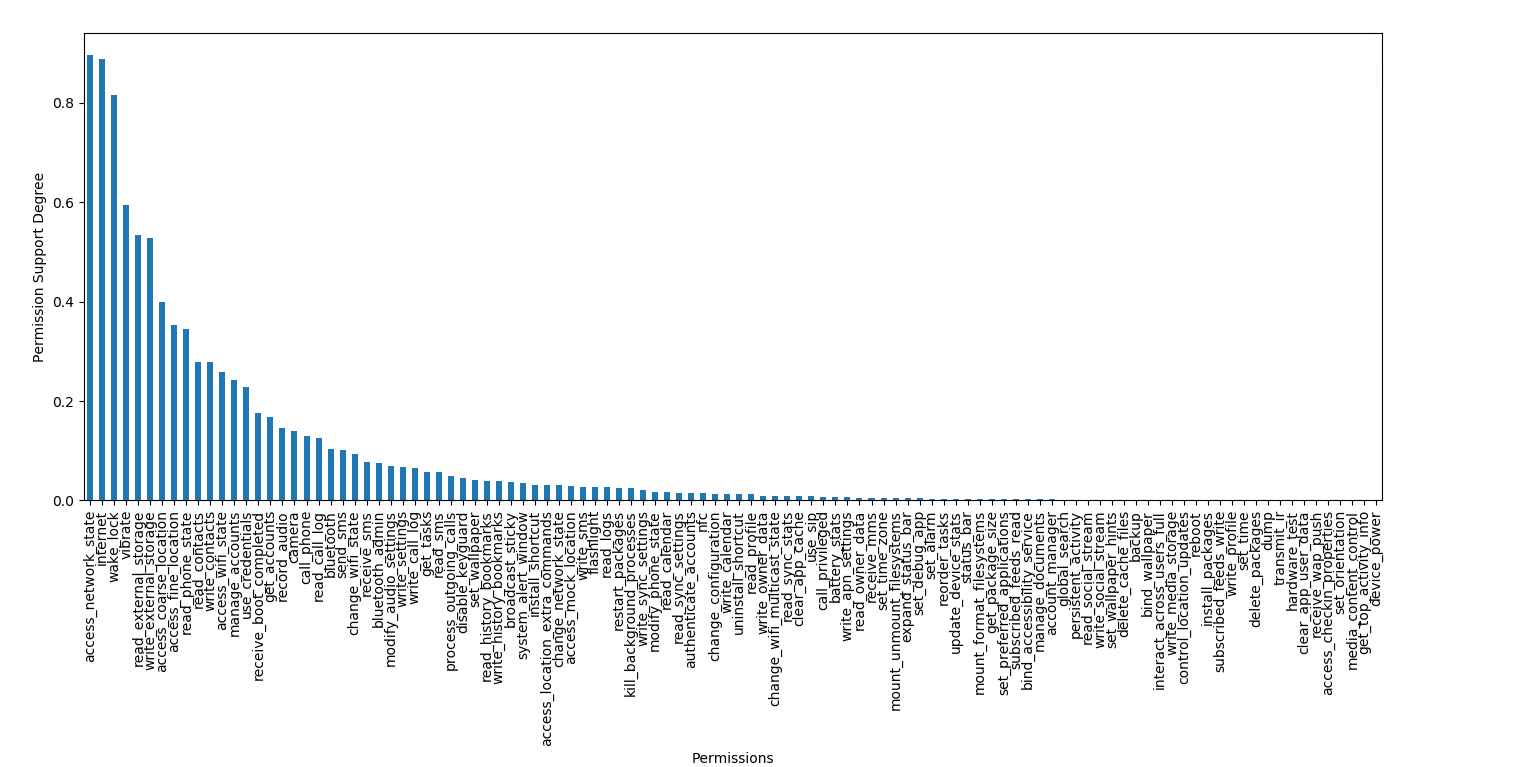} 
	\caption{{An Example of {Functionality} point-permission set identification}}
	\label{img6}
\end{figure*}
Figure 6 shows the {functionality} point-permission set relationship corresponding to the app named screener2. The permission with the highest permission support degree is permission ACCESS\_NETWORK\_STATE, and its support degree is 0.89. It is the actual permissions required by the app, and the minimum permission support degree is the permission DEVICE\_POWER, whose support degree is 0.0000146. This permission will be removed by our method since the permission support degree is too low. In this way, we have implemented the {functionality} point-permission set identification.

\subsubsection{Permission-Based Risk Evaluation}
In order to prove the effectiveness of our method, we conduct an analysis comparison on our approach and the previous method~\cite{Asb} by using the following evaluation metrics, the metrics are defined as follows:
\par
\textbf{Mean Average Precision (MAP):} $MAP$ is a comprehensive evaluation of the accuracy for the recommended permissions. In our study, to test whether the recommended permission is in the actual declared permission list, we also consider the relative order of the recommendation results, the calculation formula is as follow:

\begin{equation}
M A P=\frac{\sum_{k=1}^{M} \frac{1}{N_{k}} \sum_{l=1}^{N_{k}} \frac{R_{l}}{l} I_{l}}{M}
\end{equation}
where $M$ represents the total number of app in the test set, $N_{k}$ represents the number of permissions recommended for the $k^{t h}$ app . $R_{l}$ indicates the number of permissions that really apply for the top $l$ recommended permissions. $I_{l}$=1 indicates that the $l^{t h}$ permission in the recommended ranking is really applied by the app, otherwise $I_{l}$=0 indicates that the recommended permission does not been applied.

\textbf{The ratio of the app with Unexpected Permissions (AUPR):} The AUPR is defined as the percent of the app which has unexpected permissions in the testing dataset. 
\begin{equation}
A U P R=\frac{1}{M} \sum_{i=1}^{M} K\left(U P^{B}\left(a_{i}\right)\right)
\end{equation}
where if $\left|U P^{B}\left(a_{i}\right)\right|>0$, then $K\left(a_{i}, R i s k\right)=1$, otherwise $K\left(a_{i}, R i s k\right)=0$.

\textbf{Risk app Ratio (RAR):} The RAR is defined as the ratio of risk app in the test datasets.
\begin{equation}
R A R=\frac{1}{M} \sum_{i=1}^{M} K\left(a_{i}, \text {Risk}\right)
\end{equation}
where if $a_{i} \in \operatorname{Ris} k$, then $K\left(a_{i}, R i s k\right)=1$, Otherwise $K\left(a_{i}, R i s k\right)=0$.

\textbf{Necessary recall (NR):} The $N R$ is used to measure the recall of our approach, it defined as follow:

\begin{equation}
N R\left(A P P_{t}\right)=\frac{| A P s \text {\emph{ in top-n} } |}{n}
\end{equation}
where $n$ is the number of the necessary permissions of $A P P_{t}$, the $A P_{s}$ are $A P P_{t}$'s necessary permissions, and $\operatorname{\emph{top}}-n$ is the set of top-n permissions returned by an approach. For a set of apps, $N R$ is the mean of the $N R$ values for all apps in it.

\textbf{Total-recall ratio (TRR):} The $T R R$ is used to measure the effort our approach requires to achieve total recall, i.e., to recommend all the correct permissions for an app. It defined as follow:

\begin{equation}
\operatorname{TRR}\left(A P P_{t}\right)=\left\{\begin{array}{l}{\frac{n_{mim }}{\left|A P_{s}\right|}, \text { if } A P_{s}-R P_{s}=\emptyset} \\ {\frac{n_{a l l}}{\left|A P_{s}\right|}, \text { otherwise }}\end{array}\right.
\end{equation}
where $A P_{s}$ are $A P P_{t}$'s necessary permissions, and $R P_{s}$ are the recommend permissions by our mehtod. In a nutshell, $T R R$ measures that if we want to recall all the correct permissions of $A P P_{t}$, how many permissions on average will be recommended by an approach for one correct permission. More specifically, if the approach can achieve total recall for $A P P_{t}$, we simply compute the ratio of the minimal number of recommended permissions to achieve total recall, i.e., $n_{mim }$, 
and the number of $A P P_{t}$'s correct permissions, i.e., $\left|A P_{s}\right|$, $A P P_{t}$'s $T R R$. Otherwise, we penalize $T R R$ by replacing $\boldsymbol{n}_{mim }$ with the total number of permissions captured by the training set, i.e., $\boldsymbol{n}_{a l l}$. The closer $T R R$ is to one, the better it is. For a set of apps, $T R R$ is the mean of the $T R R$ values for all apps in it.

\par
To evaluate the performance of MPDroid, we compare it with the SF method, which is the state-of-the-art bias-based recommendation method~\cite{Asb}. In the experiments, we randomly selected 80\% of the benign app as the training set to establish a description information and permissions mapping, as well as for the malicious app. The remaining 20\% benign apps and 20\% of the malicious apps constitute the test set. After using the \emph{description-minimum permission set} identification algorithm to identify and remove over-declared permissions in benign app set, the corresponding \emph{description-minimum permission set} relationship is obtained. Next, according to the description-permission mapping, the training set is used to recommend permissions to the test apps. We set the Topic Number=100, $T_{b}$=0.6, $T_{m}$=0.4, $\theta_{\text {support }}$=0.1, and then calculate the $MAP$, $AUPR$, $RAR$ and  $ARISK$ evaluation indicators. $ARISK$ represents the average risk value for the test set app calculate by equation(16).

%
%
%
%

\begin{table}[!htbp]
	\centering
	
	\caption{Comparison of experimental results (20\% benign app as a test set)}
	\label{tab:aStrangeTable}
	
	\setlength{\tabcolsep}{1.6mm}{
		\begin{tabular}{|c|c|c|c|c|c|c|}
			\hline
			Approach & $AUPR$ &$RAR$ & $ARISK$ & $MAP$ &$NR$& $TRR$ \\  
			\hline
			\textbf{MPDroid} &\textbf{0.334}&\textbf{0.104}&	\textbf{0.335}&	\textbf{0.931}  &\textbf{0.865}&\textbf{1.257}\\
			\cline{1-7}
			SF &	0.314&	0.068&	0.200&	0.927 & 0.859& 1.489\\
			
			\cline{1-7}
			Improvement&	6.4\%&	\textbf{52.9\%}&\textbf{67.5\%}&	0.4\% & 0.6\%& 18.5\%\\

			\hline
		\end{tabular}
	}
\end{table}

\begin{table}[!htbp]
	\centering
	
	\caption{Comparison of experimental results (20\% malicious app as a test set)}
	\label{tab:aStrangeTable}
	
	\setlength{\tabcolsep}{1.6mm}{
		\begin{tabular}{|c|c|c|c|c|c|c|}
			\hline
			Approach & $AUPR$ &$RAR$ & $ARISK$ & $MAP$ &$NR$& $TRR$ \\  
			\hline
			\textbf{MPDroid} &\textbf{0.846}&\textbf{0.356}&	\textbf{2.221}&	\textbf{0.854}  &  \textbf{0.712} &	\textbf{1.72} \\
			\cline{1-7}
			SF &	0.779&	0.288&	1.817&	0.844 & 0.705& 2.619\\
			
			\cline{1-7}
			Improvement&	10.9\%&	23.6\%& 22.2\%&	0.4\% &0.1\% & 52.3\%\\
			
			\hline
		\end{tabular}
	}
\end{table}

%
%

Tables 5 and Tables 6 are the experiment results  compared with SF method. Specifically, Table 5 is a benign apps that are used as a test set. Table 6 presents the results where malicious apps are used as the test set. The experimental results shows:

\begin{itemize}
	\item 1) Under the experimental settings, MPDroid obtains higher $AUPR$, $RAR$, $ARISK$, $MAP$ and $NR$ values consistently, and $TRR$ is the opposite, which indicates higher risk identification performance.
	\item 2) Not only for the malicious apps, some of the benign apps are overprivileged (i.e., over declared the unexpected or risk permissions). In generally, the malicious apps are generally much more likely to be overprivileged than benign apps.
	\item 3) For the benign apps, the proposed MPDroid outperforms on $RAR$ and $ARISK$ compared with SF by 52.9\% and 67.5\% respectively. This shows that MPDroid can provide developers and users with more reasonable permission configuration, and reducing the over privilege problem.
	\item 4) The $MAP$ in SF and MPDroid changes little. This indicates that although the benign app has over-declared permissions, but the number of over-declared permissions is still small compared to the total number of permissions of the app. Generally, the over-declared permissions often appear in a few popular apps.
\end{itemize}

\subsubsection{Studies on the Parameters}
 In this Section, we discuss how the parameters impact the results in terms of $AUPR$, $RAR$, $ARISK$, $MAP$ since they are the main goal of MPDroid.
\subsubsection*{(1) Impact of Number of Topic }
\begin{figure*}[!t]
	\centering
	\subfigure[]{ 
		\includegraphics[width=2.25in]{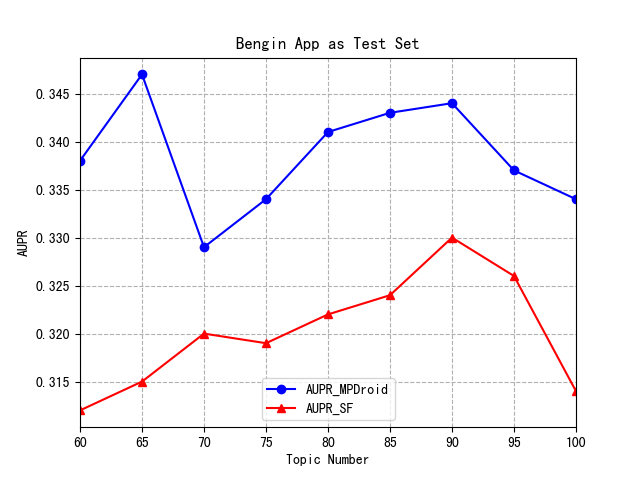} 
	} 
	\subfigure[]{ 
		\includegraphics[width=2.25in]{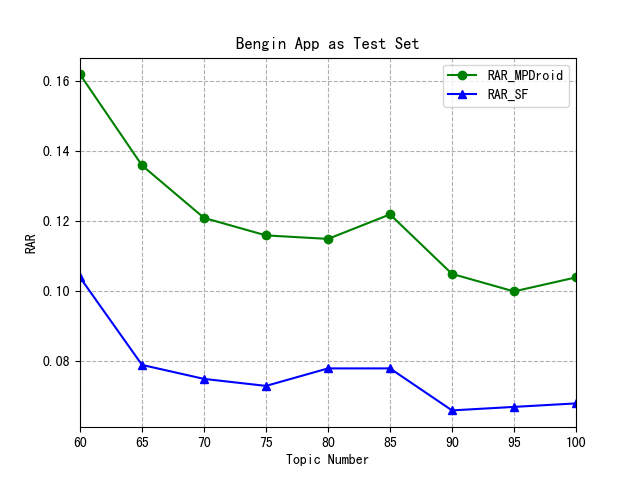} 
	}
	\subfigure[]{ 
		\includegraphics[width=2.25in]{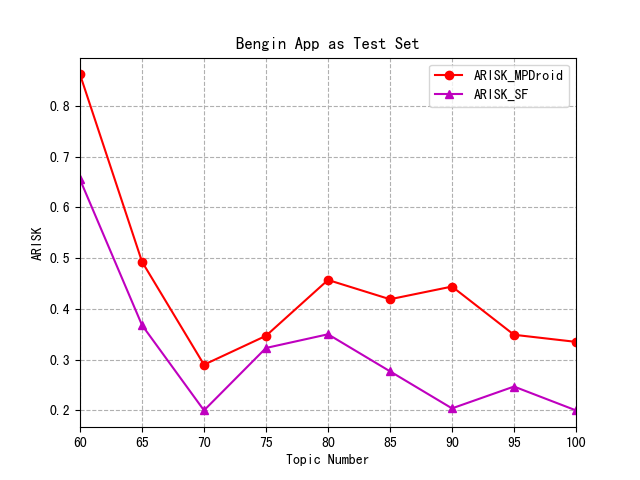} 
	}
	\subfigure[]{ 
		\includegraphics[width=2.25in]{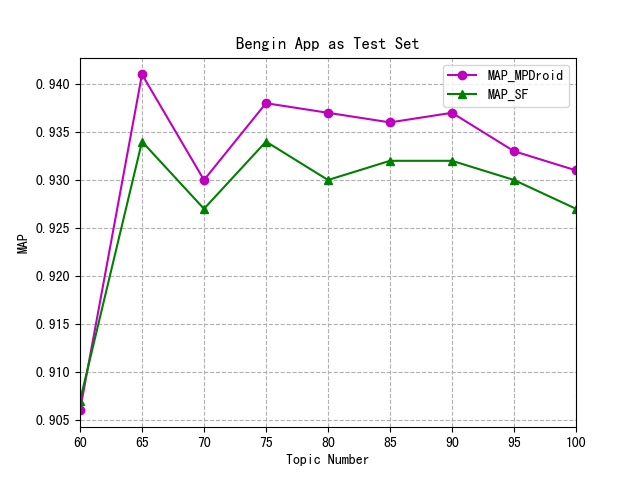} 
	}
	\caption{Benign apps as the test set}
	\label{img7}
\end{figure*}
MPDroid uses the LDA topic model to process the app description informations to obtain the functional feature vector. In order to study the impact of different topics on the final experimental results, we use the benign apps and the malicious apps as test sets respectively to evaluate the performance of MPDroid under different number of topics. In the experiments, the number of topics varies from 60 to 100 with a step value of 5, the test set ratio is set as 20\%.  In addition, according to~\cite{Asb}, we set the similarity threshold $T_{b}$=0.6, $T_{m}$=0.4 since it can obtain better results.
\begin{figure*}[!t]
	\centering
	\subfigure[]{ 
		\includegraphics[width=2.25in]{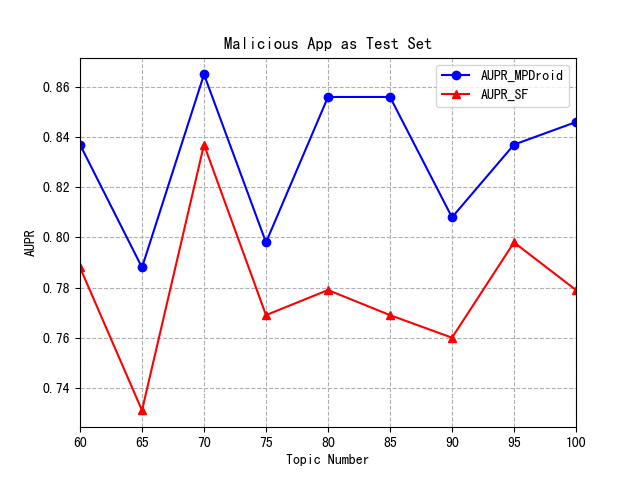} 
	} 
	\subfigure[]{ 
		\includegraphics[width=2.25in]{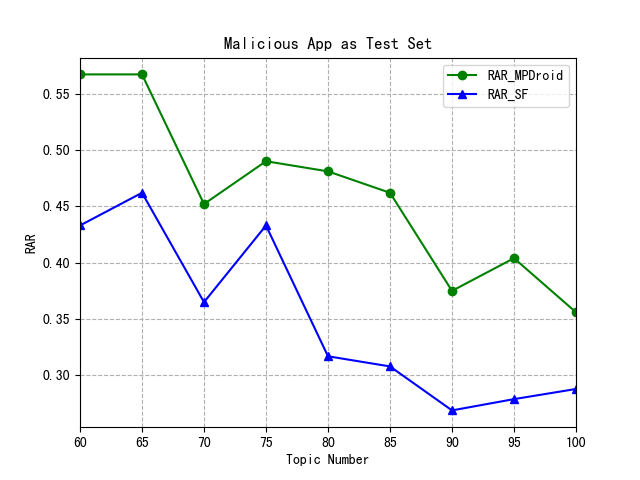} 
	}
	\subfigure[]{ 
		\includegraphics[width=2.25in]{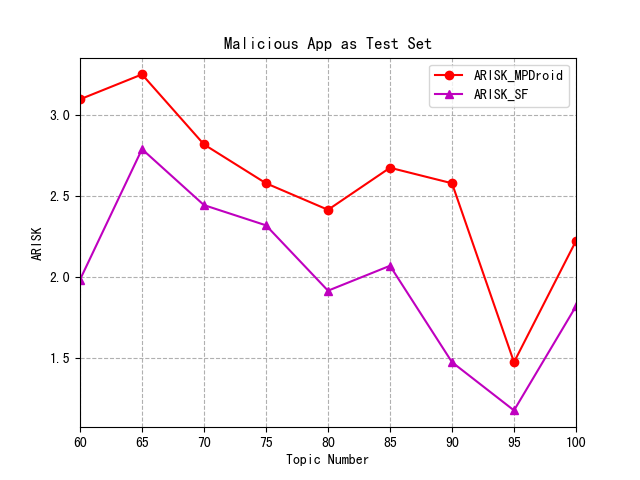} 
	}
	\subfigure[]{ 
		\includegraphics[width=2.25in]{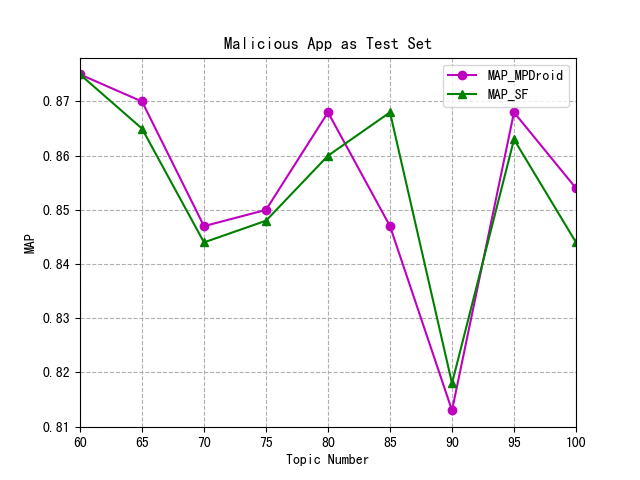} 
	}
	\caption{Malicious apps as the test set}
	\label{img8}
\end{figure*}

Figure 7 and Figure 8 show the experimental results of benign apps and malicious apps as the test set respectively, we can obtain that:
\begin{itemize}
	\item whether it is a benign app or a malicious app as a test set, MPDroid can obtain better experimental results overall. This indicates that removing the over-declaration permission of the benign app can better identify over-declared permissions.
	\item the number of topics has a certain impact on each metric. For the benign app as the test set, the $RAR$, $ARISK$ are higher when the number of topics is 60. However, for malicious app as the test set, the number of topics is 65. This indicates that the same number of topics has little impact on different test sets.
	\item For the $AUPR$ and $MAP$, the number of topics will affect $AUPR$ and $MAP$ to a certain extent, but there is no fixed rule. For example, for the benign app as a testing set, when the number of topics is 60, the $MAP$ is the lowest, and for the malicious app as the testing app set, the $MAP$ achieve the lowest value is when the number of topics is 90.
	\end{itemize}

\subsubsection*{(2) Impact of Support degree threshold $\theta_{\text {support}}$}
The support degree threshold $\theta_{\text {support}}$ in Section 3.4 mainly considers the permissions required for the actual API call in the app, which can make the recommended permissions more accurate. To study the impact of the support degree threshold, we tested the benign app and the malicious app test set separately and compared the test results with different support degree thresholds. We set the initial value is 0.0.5, and then we vary the $\theta_{\text {support}}$ from 0.1 to 0.6 with a step value of 0.1, the similarity threshold $T_{b}$=0.6, $T_{m}$=0.4, and the test set ratio is 20\%. In addition, since the support calculation only occurs in the MPDroid method, and the SF method is not affected by the parameter so that the result of SF is a horizontal line.

\begin{figure*}[!t]
	\centering
	\subfigure[]{ 
		\includegraphics[width=2.25in]{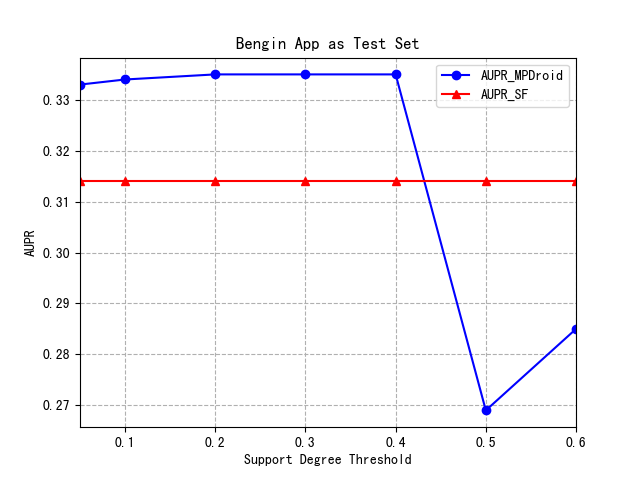} 
	} 
	\subfigure[]{ 
		\includegraphics[width=2.25in]{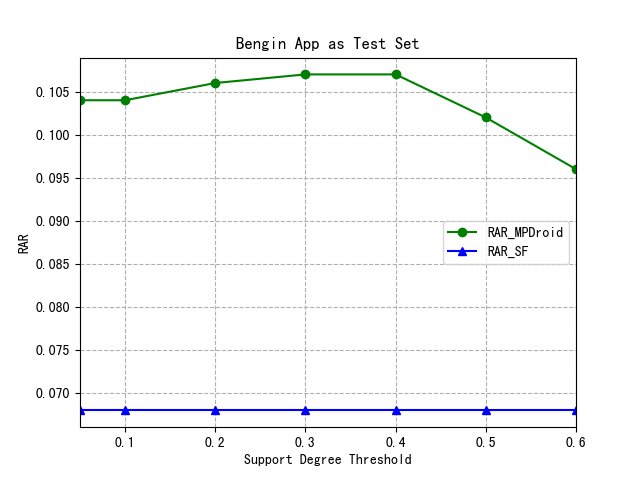} 
	}
	\subfigure[]{ 
		\includegraphics[width=2.25in]{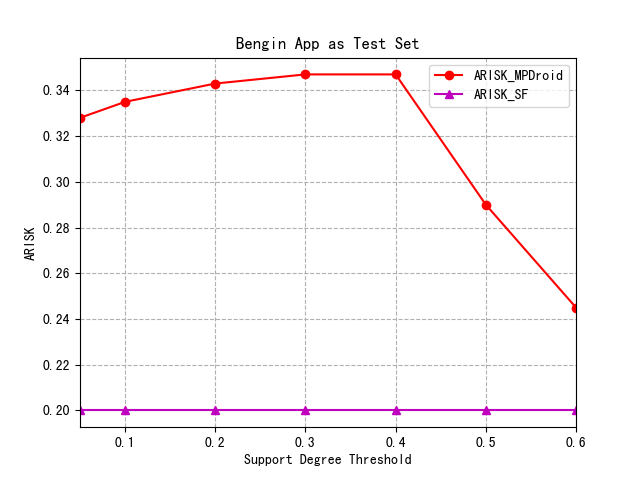} 
	}
	\subfigure[]{ 
		\includegraphics[width=2.25in]{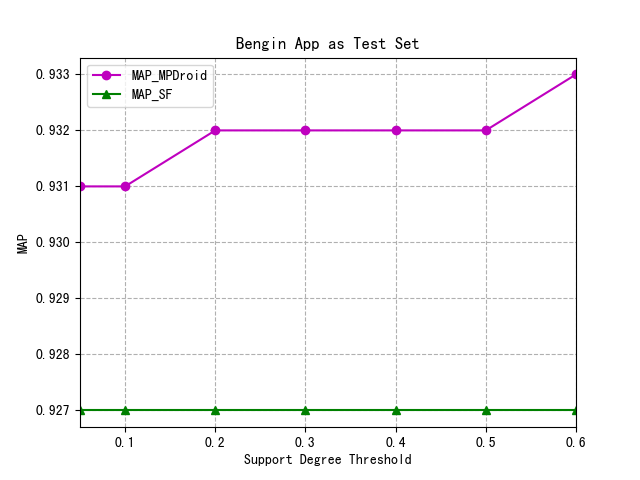} 
	}
	\caption{Benign apps as the test set}
	\label{img9}
\end{figure*}

\begin{figure*}[!t]
	\centering
	\subfigure[]{ 
		\includegraphics[width=2.25in]{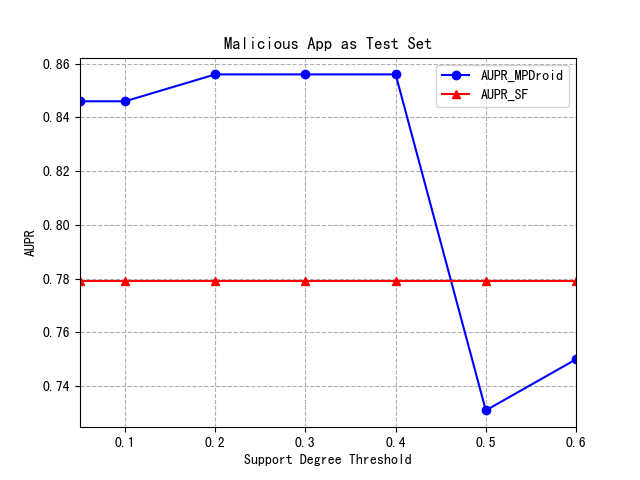} 
	} 
	\subfigure[]{ 
		\includegraphics[width=2.25in]{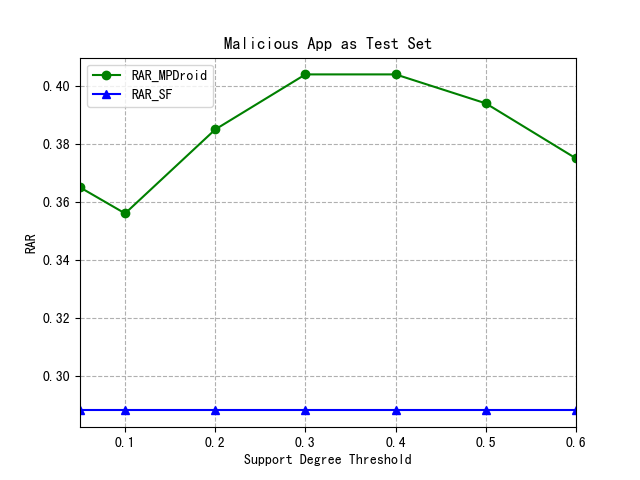} 
	}
	\subfigure[]{ 
		\includegraphics[width=2.25in]{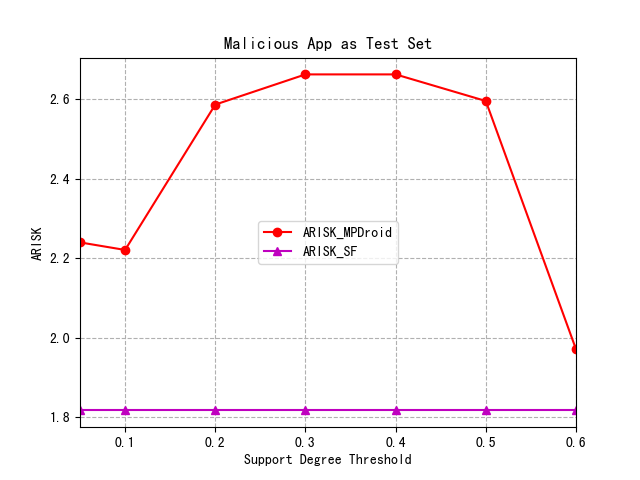} 
	}
	\subfigure[]{ 
		\includegraphics[width=2.25in]{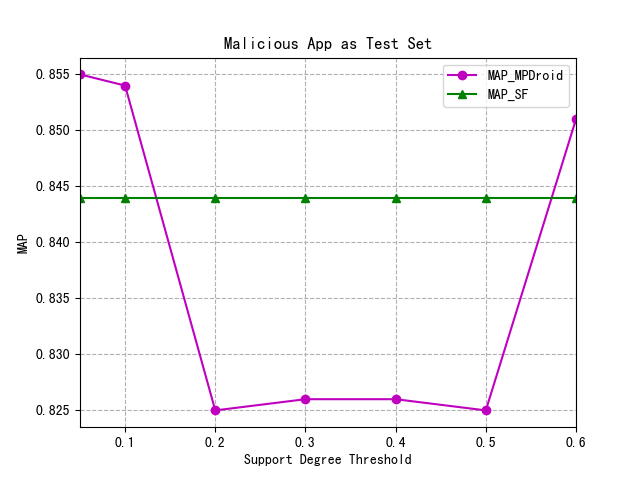} 
	}
	\caption{Malicious apps as the test set}
	\label{img10}
\end{figure*}

Figure 9 and Figure 10 show the experimental results of benign apps as the test set and malicious apps as the test set, respectively. we can know that:
\begin{itemize}
	\item Regardless of whether it is a benign or malignant test set, when the support threshold vary from 0.1 to 0.4, $AUPR$, $RAR$, and $ARISK$ change little. However, when the $\theta_{\text {support}}$ is greater than 0.4, the detection decreases, indicating that the risk permission is not filtered when the $\theta_{\text {support}}$ below 0.4. When $\theta_{\text {support}}$ is too high (greater than 0.4), the app risk permissions are filtered basically, the remaining permissions are likely to belong to the app itself and the detected risk is smaller.
	
	\item The recommendation accuracy rate of the benign test set is always greater than the malicious test set on the $MAP$. When the $\theta_{\text {support}}$ changes, the $MAP$ of the benign test set changes little, but the malicious test set changes significantly. It indicats that the risk permission contained in the malicious app are more, and it will affect the accuracy of the recommendation.	
\end{itemize}

\subsubsection*{(3)Impact of test set ratio}
The test set ratio indicates the performance of MPDroid under different data scales. In order to study the effect of different test set ratios on experimental performance, we also tested the benign apps and the malicious apps test set separately, and compared the test results under different test set ratio. In the experiments, we vary the test set ratio from 10 to 40 percent with a step value of 5 percent and the similarity threshold is $T_{b}$=0.6, $T_{m}$=0.4, $\theta_{\text {support}}$=0.1.
\begin{figure*}[!t]
	\centering
	\subfigure[]{ 
		\includegraphics[width=2.25in]{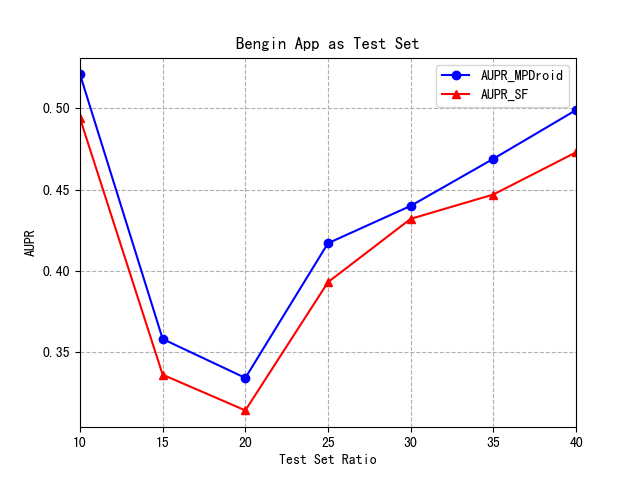} 
	} 
	\subfigure[]{ 
		\includegraphics[width=2.25in]{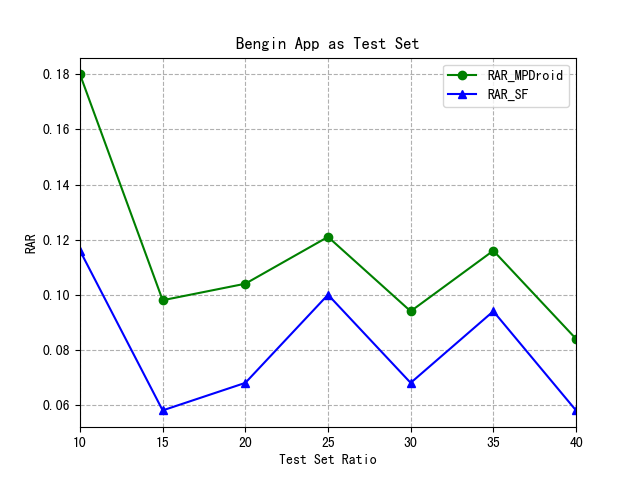} 
	}
	\subfigure[]{ 
		\includegraphics[width=2.25in]{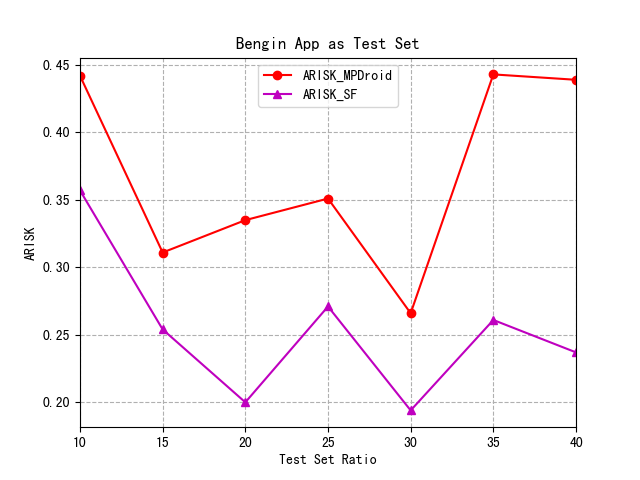} 
	}
	\subfigure[]{ 
		\includegraphics[width=2.25in]{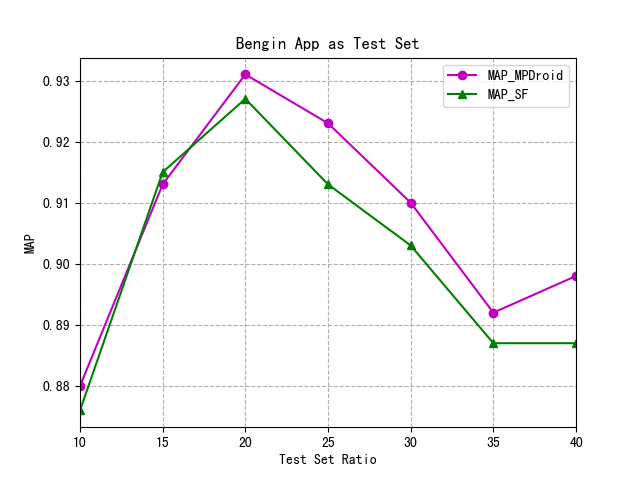} 
	}
	\caption{Benign apps as the test set}
	\label{img11}
\end{figure*}

Figure 11 and Figure 12 show the experimental results with benign apps as the test set and malicious apps as the test set, respectively. It can be obtained that:

\begin{itemize}
	
	\item For the benign app as the test set, the $AUPR$, $RAR$, $ARISK$ value are the highest when the test set ratio is 10 percent, indicating that the test performance is the best. For the malicious test set, there is no fixed rule, indicating that different data sets and test set ratios will affect the effect of the model, and MPDroid is generally better than the SF method.
		\item For the $MAP$, the recommended effect is best when the benign test set ratio is 20\%, and for the malicious test set app, the test set ratio is 25\%. It shows that different test sets and test set ratios will affect the accuracy of the recommendations.
\end{itemize}

\begin{figure*}[!t]
	\centering
	\subfigure[]{ 
		\includegraphics[width=2.25in]{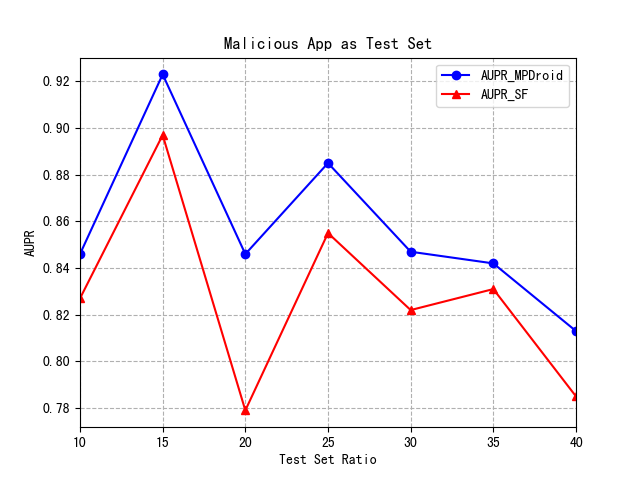} 
	} 
	\subfigure[]{ 
		\includegraphics[width=2.25in]{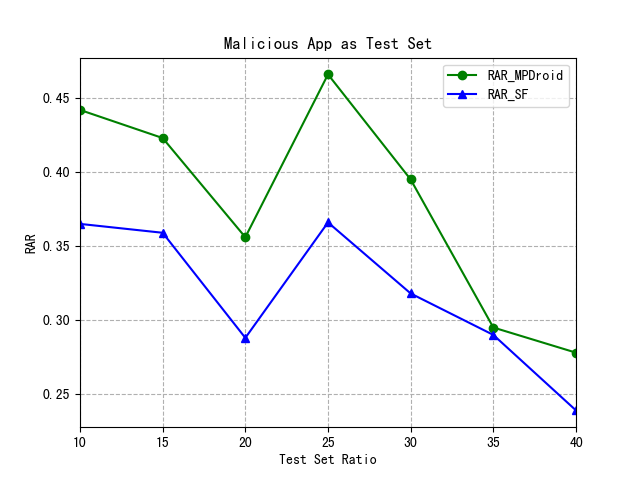} 
	}
	\subfigure[]{ 
		\includegraphics[width=2.25in]{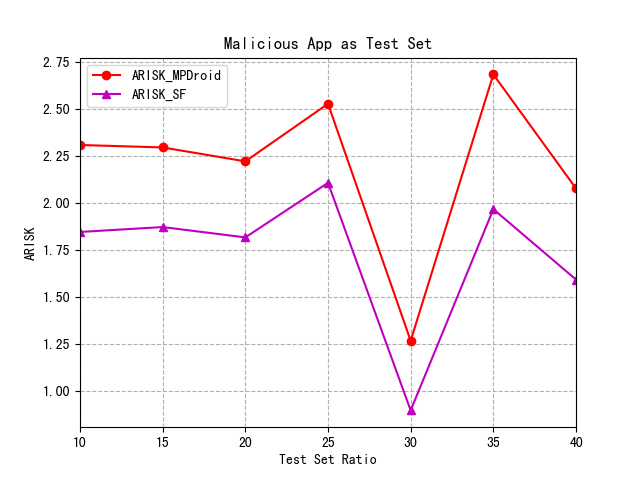} 
	}
	\subfigure[]{ 
		\includegraphics[width=2.25in]{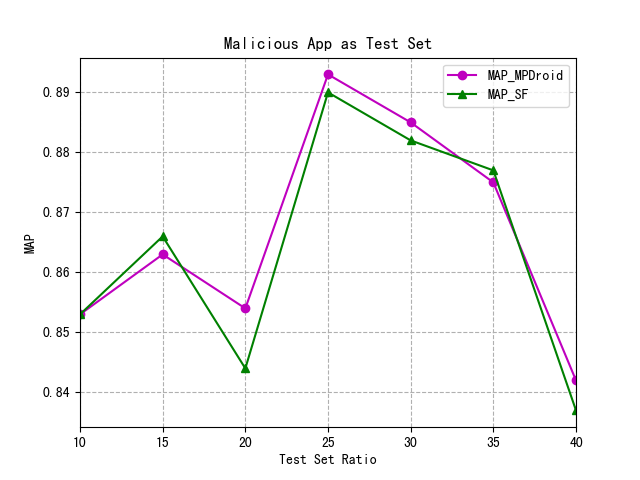} 
	}
	\caption{Malicious apps as the test set}
	\label{img12}
\end{figure*}

	\subsection{Limitations}
	MPDroid has some limitations:
\begin{enumerate}
	
	\item The implementation of MPDroid relies on the mapping the {functionalities} to permissions, but the {functionality} is only the features extracted from the app description. If the app description does not describe the {functionalities} of the app properly, the recommendation results will also be affected. To alleviate this issue, we use a large amount of app description information as a training set for training, which can largely ensure that the training model contains most of the description information, thus it can ensure the accuracy of subsequent permission recommendations. Meanwhile, we will also try to extract more granular and accurate functional features, such as adding other auxiliary information in future. In addition, we would add manual annotation if necessary.
	
	\item Currently, the method of \emph{description-minimum permission set} identification relies on the recommendation precision, and the establishment of model depends on data set. To alleviate this issue when we choose the dataset, the apps with at least 100+ downloads and five stars in Google Play are considered as the benign dataset, and the malicious app set we select from VirusShare website directly, which can slove the problem to some extent. In the future, we will enrich the mechanism for model building and further optimize the model to make it more generalized.
	
	\item In the static analysis phase of the model, we cannot achieve the completely correct permissions to parse the app since some malicious apps re-packaging, or the interference of code mixing. Generally, some developers are not familiar with Android's business logic and lead to some useless code, result in matching too many permissions. Therefore, we will focus on improving static analysis and increasing the accuracy of the analysis in our follow-up work.
\end{enumerate}

\section{RELATED WORK}
 The spurt development of mobile Internet has promoted the security problem of mobile app to become a hot spot in the industry. Due to the imperfection of the Android permission mechanism, some irresponsible developers use the permissions of the Android app indiscriminately, and the user does not realize the importance of the problem so that led to a series of privacy leaks and would pose significant risks to users and the entire mobile ecosystem.
	\par
One cornerstone of the Android security design is the permission system. An app must hold the permissions to successfully access the security and privacy critical methods. The permissions or the risk caused by the permissions have received a lot of attentions by many security types of research. Some studies are also dedicated to reminding users to potential risks through study permission dialogs, further or through improvements to enhance the permissions dialog~\cite{ApU}~\cite{Rtd}, but the impact of these methods on end users is still very small, users may still experience the risk behind it.
	\par
	The above research is mainly based on the permission mechanism already contained in Android, and does not modify the underlying architecture of Android permissions. Their research content mainly focuses on two aspects. On the one hand, it is aimed at users, they try to make users understand the permissions by giving hints. For example, Kelley~\cite{Pap} designs the dialog content of the permission prompt box in more detail, reminds the user of the consequences when using specific permissions and what is the implied risk will cause, but they does not explain the details of the resources accessed and whether it may lead to the disclosure of some private information, Bin Liu ~\cite{Fmr} propose a methodology PPA to learn privacy profiles for permission settings and leverage these profiles in a personalized privacy assistant that actively supports users in configuring their permission settings. On the other hand, it is mainly facing for developers, it remind the developers to strengthen their understanding of permissions, do not declare unnecessary or wrong permissions during the development process, but because Android development document do not have mature API descriptions, not only not enough details, but there may be some errors~\cite{HtA}. So in a summarize, no matter whether it is for users or developers, the issue of permissions has always been an urgent problem to be solved.
	\par
	In order to solve the above problems, many studies decided to turn to use machine learning~\cite{Experimental}~\cite{Par} or  recommended algorithms to solve the problem of permission redundancy, and then recommend reasonable permissions. For example, Min Peng~\cite{Par} proposed a app risk score calculating method ARSM based on app-permission bipartite graph model which combine the correlation of apps’ permissions and users’ interests. Whyper~\cite{TAR} and AutoCog~\cite{AMt} extracts various feature word combinations from the description information of the Android app as the app's declaration {functionality} and map them to the app declared permission, and through the gap between them to determine whether the declared permission meets the described {functionality}, and further determine the risk.
	\par
	Besides, there are some other methods to extract a variety of feature vectors through code analysis to detect malicious apps. For example, Bartel~\cite{Asp} proposed a tool named ``COPES" to detecting permission gaps using static analysis. It extracts from the Android framework bytecode a table that maps every method of the API to a set of permissions the method needs to be executed properly. Mujahid~\cite{SPR} implement an technique in a tool called PERMLYZER which automatically detects permission issues from apps APK to study the permission related issues in Wearable apps. More recently, some studies~\cite{Maa}~\cite{Wps}~\cite{Aaa} also focus on recommend permissions by the used APIS, such as Karim~\cite{Maa} presented a tool named ApMiner which combines static analysis and association rule discovery to make app permission recommendations, and the results shown that ApMiner performs better than PScout and Androguard in terms of app permission recommendations. However, the average F1-score of APMiner is not sufficiently high (only approximately 55\%) for it to be used in practice. Bao~\cite{Wps} propose an approach named APRecCF ,which is based on collaborative filtering technique, the start point is that apps that use similar APIs often support similar features, it measures the similarity of two apps based on the APIs used by the apps and the result show that their approach achieve significant improvement in terms of the precision, recall, F1-score and MAP of the top-k results over Karim et al.’s approach. More over, based on ~\cite{Wps}, Bao~\cite{Aaa} also propose two novel approaches to realize permission recommendation, the first approach utilizes a collaborative filtering technique utilizes and the second approach recommends permissions based on a text mining technique that uses a naive Bayes multinomial classification algorithm, which show a better preference than the others. To best of our knowledge, the best work of permission recommendation is PerRec~\cite{Ahc}, which leverages mining-based techniques and data fusion methods to recommend permissions for given apps according to their used APIs and API descriptions, and their results show that PerRec significantly improves the state-of-the-art approaches APRecCF~\cite{Aaa}, APRecTEXT~\cite{Aaa} and Axplorer~\cite{Odt}.
	
		\par
	The main disadvantage of the above methods is that the permissions declared by the app are treated as actually needed permissions by the app, but this is not always true, Android apps tend to be overprivileged. According to several studies, e.g.,~\cite{Apd} and~\cite{Rtd}, overprivileged apps are not the minority, but the majority. Therefore, it is necessary to identify the permissions that an app really needs, that is, its minimum permissions set as the permissions the app actually needs. Further, {if given a {functionality point}}, we can find a way to identify the minimum permission set corresponding to the {functionality} point, i.e., an app can declare the permissions to implement this {functionality} point. Then, it is possible to significantly improve the performance of over-declared permission detection so that to find a minimum set of permissions for the app of its describe information.
	\par
	In order to achieve the goals, different from the above methods, at a higher level, our study is similar and creative to these former approaches. Our approach combines static analysis and collaborative filtering to identify the minimum permission set of Android apps. We measure the textual descriptions of the app in Google Play and the relevant permissions it required in a more accurate way.

\section{CONCLUSION AND FUTUREWORK}

Detecting the overprivileged permissions for the mobile applications is considered as a critical and valuable task to enhance the permission system for the mobile service ecosystem. Considering the miss match between the declared {functionalities} and the requested permissions to support the declared {functionalities}, we propose an iteration approach that combines static analysis and collaborative filtering to identify the minimum permission set for the mobile apps. The static analysis is used to achieve the real requested permissions for each mobile app, and a \emph{description-minimum permission set} iteration algorithm based on collaborative filtering is developed to mine the relationships between the declared {functionalities} and the minimum requested permissions, so that we can detect the over-requested permissions and identify the high risk applications. Comparing with the previous state-of-the-art method, our MPDroid method can effectively identify the abnormal use of permissions and generate better permission recommendation configuration results, thereby reducing the problem of mismatch between {functionalities} and permissions. For the users, only need to have a description information of the app, so we can predict the permission required for the description information according to our model. In addition, this model also can be used to analyze the permissions of existing Android applications and evaluate the risks of the app.
\par
For the future works, we will consider more declarative {functionality} information, such as user's comment information, app privacy policy information, and the  classification information to improve the accuracy of the description {functionality}. At the same time, we will also add a new data cleaning mechanism to ensure the security of the declare permissions. In addition, since the \emph{description-minimum permission set} identification module relies on datasets, we will further improve our model by focusing on related data cleaning and mining techniques, combining with the most advanced permission identification algorithms. We envision our method would help the developers to configure the requested permissions and design the declared {functionality}.

\section*{Acknowledgment}

This work is supported by the National Key R\&D Program of China grant No.2017YFB1401201, the National Natural Science Foundation of China grant No.61572350, the National Natural Science Key Foundation of China grant No.61832014 and the Shenzhen Science and Technology Foundation (JCYJ201708\\16093943197).

\section*{References}

\bibliography{reference}

\end{document}